\documentclass[aps,prl,onecolumn,showpacs]{revtex4}
\usepackage{amssymb}
\usepackage{graphicx}
\usepackage{epsfig}

\input{tcilatex}

\begin{document}

\date{\today}
\title{The glass transition and the Coulomb gap in electron glasses}
\author{M.\ M\"uller, L. B. Ioffe}
\affiliation{Department of Physics, Rutgers University, Piscataway, New Jersey 08854 }

\begin{abstract}
We establish the connection between the presence of a glass phase and the
appearance of a Coulomb gap in disordered materials with strongly
interacting electrons. Treating multiparticle correlations in a systematic
way, we show that in the case of strong disorder a continuous glass
transition takes place whose Landau expansion is identical to that of the
Sherrington-Kirkpatrick spin glass. We show that the marginal stability of
the glass phase controls the physics of these systems: it results in slow
dynamics and leads to the formation of a Coulomb gap.
\end{abstract}

\maketitle

The relation between the slow dynamics of Coulomb glasses, disordered
materials with strong electron-electron repulsions, and the appearance of a
soft "Coulomb" gap in their density of states (DOS) has been a mystery for a
long time. The strong effect of interactions on the DOS was first noticed by
Pollak \cite{Pollak70} and Srinivasan \cite{Srinivasan71}. Efros and
Shklovskii \cite{efrosshklovskii7576} predicted the Coulomb gap as resulting
from the long-range Coulomb interactions between localized electrons in
semiconductors, and leading to a crossover in the temperature dependence of
the conductivity from Mott's law $\ln (\rho )\sim (T_{M}/T)^{1/4}$ to $\ln
(\rho )\sim (T_{ES}/T)^{1/2}$~\ at low temperatures \cite{ESbook}. The
Efros-Shklovskii law for the conductivity was verified in semiconductors,
alloys and granular metals, and recently the gap itself was directly
observed in semiconductors~\cite{Cbgap,Sandow}. The presence of disorder
frustrates the Coulomb interactions and leads to glassy behavior in such
materials, as predicted theoretically long ago~\cite{Cbglass}. The first
evidence for glassy phenomena came from the slow relaxation of charge
injected into compensated semiconductors ~\cite{DonMonroe87}. Later,
Ovadyahu's group established that the slow dynamics in Indium-oxides is
indeed due to electron-electron interactions~\cite{EGnonergodic}. Very
recently the same group has demonstrated memory and aging effects similar to
those observed in spin glasses ~\cite{EGaging}.

Despite the experimental progress a thorough understanding of the glass
phase is still missing. The source of the difficulty is that in order to
describe glassy phenomena one needs to take electron correlations into
account, while the approach by Efros and Shklovskii is essentially a single
particle theory. The necessity to include correlations has become clear from
several recent numerical studies of the off-equilibrium dynamics of Coulomb
glasses~\cite{EGdynnum}. Further, an increasing number of experiments
suggests that a quantitative description of the hopping conductivity should
take multiparticle processes and correlations into account~\cite%
{multiparticle}. One also needs to go beyond a single particle theory in
order to understand the relation between the Coulomb gap and the glass
transition. The mean field solution of a model of uniformly interacting
electrons in a disordered medium indicates that the glass transition and the
formation of a pseudogap in the DOS are driven by the same mechanism, and a
similar relation has been conjectured for the Coulomb glass~\cite{pastor99}.

The goal of this paper is to develop a formalism that accounts for the
correlations between the electrons in a realistic model for Coulomb glasses
in 3D. Our approach is based on the locator approximation developed for spin
glasses in Refs.~\cite{Feigelman79, BrayMoore79,LopatinIoffe02}. The idea is
to map the original lattice problem into an effective single-site problem
that encodes correlations by the distribution of a fluctuating local field,
which gives exact results for infinite range models. In the limit of strong
disorder, the Coulomb interactions are essentially unscreened, so that the
effective number of neighbors is large and the locator approximation is
parametrically well justified. In this regime, we find a replica symmetry
breaking glass transition, which belongs to the same universality class as
the transition in the Sherrington-Kirkpatrick (SK) spin glass~\cite%
{BrayMoore78}. Below the transition, the phase space divides into an
exponential number of metastable states and ergodicity is broken. Like any
generic glass, this electronic system freezes into one of many marginally
stable states since the latter are the most abundant (the number of states
increasing exponentially with decreasing stability). Marginally stable
states are characterized by the presence of soft modes that lead to the slow
relaxation dynamics observed in experiments. Above $T_{c}$, the DOS does not
display any particular signature of the Coulomb interactions. We show that
the Coulomb gap forms only below $T_{c}$ where it emerges as a direct
consequence of marginal stability. Finally, we derive an asymptotic
expression for the DOS at very low temperatures.

We consider the classical model~\cite{ESbook} for strongly localized
electrons occupying a fraction $0<K<1$ of a given set of impurity sites $i$, 
\begin{equation}
H=\frac{1}{2}\sum_{i\neq j}n_{i}\mathcal{J}_{ij}n_{j}+\sum_{i}n_{i}(\epsilon
_{i}-\mu _{K}),  \label{Hamiltonian}
\end{equation}%
where $n_{i}\in \{0,1\}$ is the occupation number of the site $i$. For
simplicity, we take them to be arranged on a cubic lattice with lattice
spacing $\ell \equiv 1$. The unscreened Coulomb interactions are described
by $\mathcal{J}_{ij}= 1/r_{ij}$ in units where $e^{2}/\ell \equiv1$. The $%
\epsilon _{i}$'s denote random on-site energies, and $\mu _{K}$ is the
chemical potential. We restrict ourselves to the case $K=1/2$ where the
particle-hole symmetry implies $\mu _{1/2}=0$, and suggests to introduce
spin variables $s_{i}=n_{i}-1/2=\pm 1/2$. Further, we assume a Gaussian
distribution of width $W$ for the on-site energies $\epsilon _{i}$. Their
randomness emulates the effect of the disorder in the site positions which
is present in all physical electron glasses and generates rather large
site-to-site fluctuations of the Coulomb potential. We focus on the limit of
strong disorder, $W\gg 1$ and dimension $D=3$. In this case screening is
suppressed on short scales and the interactions remain long-range, which
justifies the use of the locator approximation. We will see that at low
temperatures the self-generated disorder outweighs $W$. We thus expect our
results to be universal in that regime. Furthermore, this observation makes
us believe that at low temperatures the locator approximation can be
justified even in the case of weak disorder.

\begin{figure}[tbp]
\resizebox{7.5 cm}{!}{\includegraphics{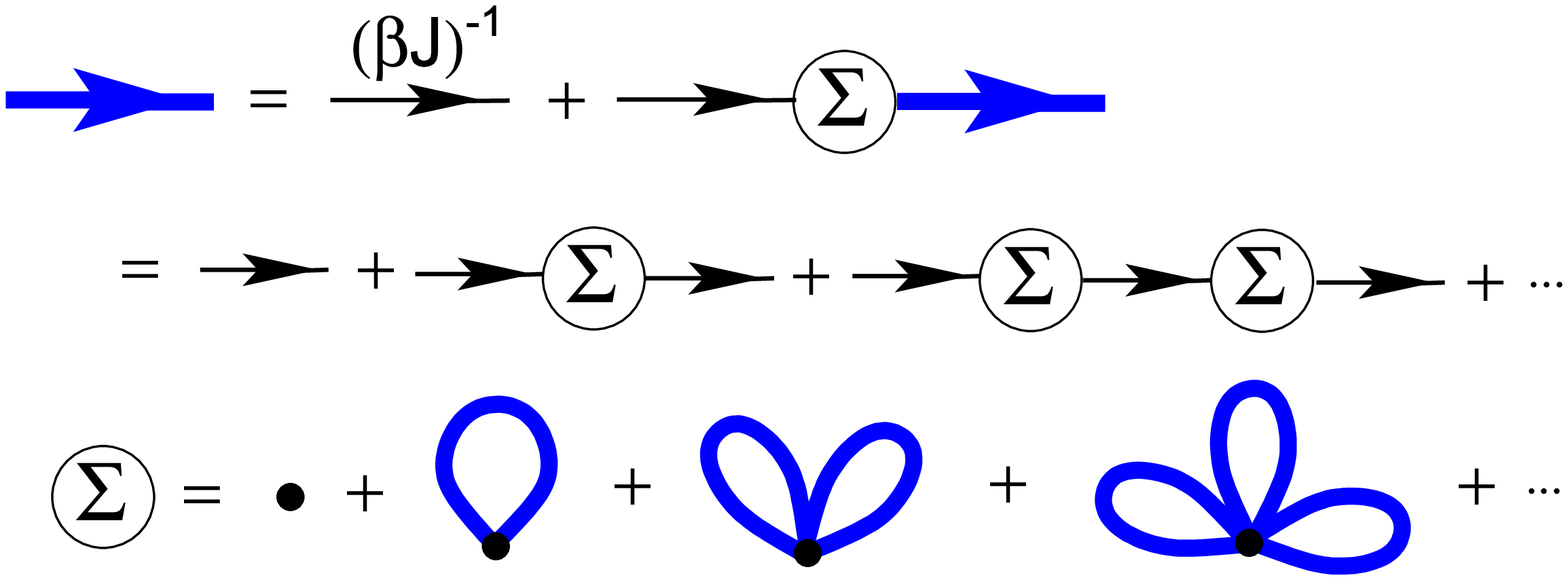}}
\caption{For long-range interactions the self-energy $\Sigma$ can be
approximated by a local operator. The full propagator is obtained as a
simple geometric series.}
\label{locator}
\end{figure}
In the case of long-range interactions, diagrammatic expansions can be
efficiently resummed since the large number of effective neighbors allows
one to approximate the self-energy by an average local term, see Fig.~\ref%
{locator}. This observation further suggests to replace the interactions of
a given spin with its environment by an effective local field described by
couplings of the spin to its own replicas. This reduces the model to a
single-site problem, translating the complexity of the environment into a
non-trivial replica structure of the one-site Hamiltonian~\cite%
{LopatinIoffe02,pastor99}, 
\begin{equation}
\beta H_{0}(\{s_{\alpha }\})=\frac{1}{2}\sum_{a,b}s_{a}\mathcal{B}%
_{ab}s_{b}+\beta \sum_{a}s_{a}\epsilon _{0}.  \label{H0}
\end{equation}

We note that this procedure resembles the way in which the SK-model is
transformed into a one-site problem. Indeed, the locator approximation
applied to the SK-model sums all tree-like diagrams with doubled interaction
lines which becomes exact in the large $N$ limit.

To make contact between the Hamiltonians (\ref{Hamiltonian}) and (\ref{H0}),
we require that they both yield the same single-site spin correlation
functions, 
\begin{eqnarray}
\left\langle s_{a}s_{b}\right\rangle =\sum_{\{s_{\alpha }\}}s_{a}s_{b}e^{-%
\frac{1}{2}\sum_{\alpha ,\gamma }s_{\alpha }\tilde{\mathcal{B}}_{\alpha
\gamma }s_{\gamma }}=\left[ \frac{1}{\tilde{\mathcal{B}}-\Sigma }\right]
_{ab}  \label{SP} \\
=\frac{1}{N} \sum_{i}\left\langle s_{i,a}s_{i,b}\right\rangle =\frac{1}{V}Tr%
\left[ \frac{1}{\beta \mathcal{J}-\beta ^{2}W^{2}\mathcal{I}-\Sigma }\right]
_{ab}.  \nonumber
\end{eqnarray}%
We have averaged over the random energies $\epsilon _{i}$, and as usual, the
number of replicas $n$ is implicitly assumed to tend to zero in the end of
calculations. $\mathcal{I}$ denotes a $n\times n$ block matrix with all
entries equal to 1, and we have defined $\tilde{\mathcal{B}}=\mathcal{B}%
-\beta ^{2}W^{2}\mathcal{I}$. In Eqs.~(\ref{SP}) we approximated the full
propagator for either model as a simple geometric series with a \emph{local}
self-energy insertion $\Sigma _{ab}\delta _{ij}$, as motivated above. Since
the mapping is to preserve correlations, the self-energy has to be the same
for both models~\cite{LopatinIoffe02}. From (\ref{H0}), we obtain the free
energy 
\begin{eqnarray}
n\beta F(\mathcal{B}) = -\ln \left[ \sum_{s_{\alpha }}e^{-\frac{1}{2}%
\sum_{\alpha ,\gamma }s_{\alpha }\tilde{\mathcal{B}}_{\alpha \gamma
}s_{\gamma }}\right] +\frac{U(\mathcal{B})}{2}  \label{F(B)}
\end{eqnarray}
where $U(\mathcal{B})$ has to be determined such that the saddle point
equations with respect to $\mathcal{B}$ yield back Eqs.~(\ref{SP}). Up to a
function of temperature, we find 
\begin{eqnarray}
U(\mathcal{B}) = tr\left[ \ln (\tilde{\mathcal{B}}-\Sigma )- \frac{1}{V}%
Tr\ln \left( \beta \mathcal{J}-\beta^{2}W^{2}\mathcal{I}-\Sigma \right)%
\right]  \label{U(B)}
\end{eqnarray}%
where $tr$ denotes the trace in replica space. We emphasize that in this
expression the self-energy $\Sigma $ has to be considered as an implicit
function of $\mathcal{B}$ as defined through Eq.~(\ref{SP}). In the
following we shall need spatial traces like $g_{n}(x)=V^{-1}Tr\frac{1}{%
(\beta \mathcal{J}+x)^{n}}$ which we evaluate in Fourier space $%
g_{n}(x)=\int $ $\frac{d^{3}k}{(2\pi )^{3}}\frac{1}{(\beta J_k+x)^{n}}$ with 
$J_{k}=4\pi /k^{2}$ at small $k$. We assume some cut-off procedure that
regularizes the small scale physics so that $\int_{k}=1$ and $%
\int_{k}J_{k}=0 $. For $x\gg \beta$ we obtain $g_{n}(x)\approx
x^{-n}[1+C_{n}(\beta/x) ^{3/2}]$ where $C_{1}=2\sqrt{\pi }$ and $C_{2}=5%
\sqrt{\pi }$.

Let us first discuss the replica symmetric (RS) solution of Eqs.~(\ref{SP})
for which we assume $\Sigma _{ab}=-\Sigma _{0}\delta _{ab}+\Sigma _{\mathcal{%
I}}\mathcal{I}$, and $\mathcal{B}_{ab}=-B_{0}\delta _{ab}+B_{\mathcal{I}}%
\mathcal{I}$. For $W\gg 1$, we find $\Sigma _{0}\approx \sqrt{2\pi }\beta W$%
, suggesting the interpretation of $\Sigma _{0}^{-1}$ as the fraction of
thermally active sites (for $T_{c}<T\ll W$). The distribution of local
fields obtained from this RS solution matches remarkably well numerical
simulation data for finite temperatures and $W\lesssim 1$, even though the
locator approximation is difficult to justify in this parameter regime~\cite%
{pankov04}. A depletion of sites in small fields is found due to strong
correlations in this \textquotedblleft Coulomb plasma\textquotedblright .
However, a closer analysis reveals that there is no true pseudogap on the
replica symmetric level, the depletion disappearing completely for strong
disorder. This is also reflected in the charge susceptibility $\chi _{%
\mathrm{RS}}=\beta (1/4-\left\langle s_{a}s_{b}\right\rangle _{a\neq
b})\approx \beta /\Sigma _{0}$ which tends to a finite constant ($\sim 1/W$)
within this solution. The genuine Coulomb gap is formed only when the
replica symmetry is broken. For $W\gg 1$, the RS solution indeed exhibits an
instability when the condition 
\begin{eqnarray}
&&\int_{-\infty }^{\infty }dy\frac{e^{-y^{2}/2(W^{2}+B_{\mathcal{I}}/\beta
^{2})}}{\sqrt{2\pi (W^{2}+B_{\mathcal{I}}/\beta ^{2})}}\frac{1}{[2\cosh
\left( \beta y/2\right)]^4 }  \label{RSinstab} \\
&&\quad =\left[ g_{1}^{-2}(\Sigma _{0})-g_{2}^{-1}(\Sigma _{0})\right]
^{-1}\approx (\pi \beta ^{3}\Sigma _{0})^{-1/2},  \nonumber
\end{eqnarray}%
is met, from which we extract the critical temperature $T_{c}\approx
W^{-1/2}/[6(2/\pi )^{1/4}]\ll 1\ll W$. We emphasize that the difference $%
g_{1}^{-2}(\Sigma _{0})-g_{2}^{-1}(\Sigma _{0})$ is controlled by the
contribution from large scales, $\ 1/k\sim \sqrt{W}$, which justifies our
assumption of a large number of effective neighbors.

The instability (\ref{RSinstab}) signals a continuous glass transition with
full replica symmetry breaking. We may analyze it further by expanding the
free energy with respect to the replicon mode $\delta \mathcal{B}$ (with $%
\delta \mathcal{B}_{\alpha \alpha }=0$ and $\delta \mathcal{B}\mathcal{I}=0$%
) around the RS solution, 
\begin{equation}
n\beta \delta F=\frac{c_1}{W^{3/2}}\left[ tr(-\tau \delta \mathcal{B}^{2}+
c_{2} \delta \mathcal{B}^{3}) +c_{3} \sum_{\alpha ,\gamma }\delta \mathcal{B}%
_{\alpha \gamma }^{4}\right]
\end{equation}%
where $\tau =1-T/T_{c}$. This shows that the glass transition in Coulomb
glasses belongs to the same universality class as the one in the SK-model.
Hence, many results about the critical behavior known for infinite range
spin glasses~\cite{CugliandoloKurchan9394} should be directly applicable to
the present case. This might be interesting in particular for the aging and
memory effects observed in experiments \cite{EGaging}, even though the
dynamics of spin glasses obey slightly different rules than in Coulomb
glasses. Vice versa, electron glasses present an appealing testing ground
for many theoretical ideas developed in the context of the SK-model.

\begin{figure}[tbp]
\resizebox{7.5 cm}{!}{\includegraphics{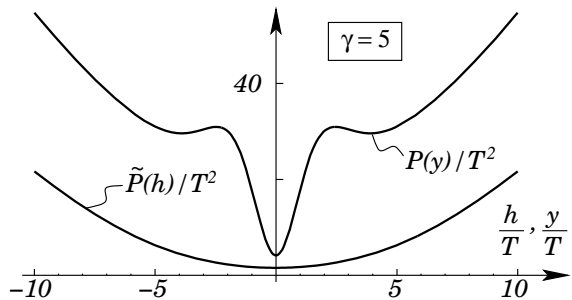}}
\caption{The distributions $\widetilde{P}(h)$ and $P(y)$ of the local and
thermodynamic fields, respectively. Since the latter allow for relaxation of
the environment, the gap is much narrower.}
\label{fielddist}
\end{figure}

We now turn to a more detailed analysis of the physics far below $T_{c}$.
Since we expect that $\Sigma _{0}\approx \beta \chi ^{-1}\gg \beta $ we may
expand the free energy (\ref{U(B)}) in $\beta \mathcal{J}/\Sigma $. Using
Eqs.~(\ref{SP}) we eliminate $\Sigma $ and obtain $U(\mathcal{B})=-tr(%
\mathcal{B}^{3})/12\pi \beta ^{3}$, resembling the SK-model where $U(%
\mathcal{B})\sim -tr(\mathcal{B}^{2})/\beta ^{2}$. The exponent reflects the
spatial dimension $D=3$ and is responsible for the shape of the pseudogap ($%
\rho (E)\sim E^{D-1}$). In order to derive this result, it is more
convenient to keep the self-energy in the formalism. Let us suppose that the
replica symmetry is broken at the level of $K$ steps. We represent the
Parisi matrices as $\Sigma =-\Sigma _{0}+\sum_{k=1}^{K}\Sigma _{k}R_{m_{k}},$
where $R_{m_{k}}$ consist of blocks of size $m_{k}$ on the diagonal with all
entries equal to 1. Let us focus on the set $\mathcal{C}$ of the $m_{1}$
spins corresponding to one of the innermost blocks. These spins experience
an effective field $y$ created by all other spins. We describe its thermal
fluctuations by a distribution $P(y)$, which in the RS case was a simple
Gaussian (see Eq.~(\ref{RSinstab})). In the case of continuous replica
symmetry breaking, $P(y)$ can in principle be obtained by integration of
Parisi's differential equation using the methods of Refs.~\cite%
{SDtechnique84}. Here, we will only exploit the fact that the Coulomb glass
is in a marginally stable state (the Hessian $\partial ^{2}F/\partial 
\mathcal{B}^{2}$ has a vanishing eigenvalue in the replicon mode $\delta 
\mathcal{B}$ characterized by $\delta \mathcal{B}_{\alpha \alpha }=0$ and $%
\delta \mathcal{B}R_{m_{k}}=0$ for all $k$), which imposes the constraint 
\begin{equation}
\int_{-\infty }^{\infty }dyP(y)\frac{1}{[2\cosh (\beta y/2)]^{4}}=\frac{1}{%
g_{1}^{-2}(\Sigma _{0})-g_{2}^{-1}(\Sigma _{0})}.  \label{margstab}
\end{equation}%
Further, the innermost component of Eqs.~(\ref{SP}) reads 
\begin{eqnarray}
\chi &\equiv &\beta \left[ \frac{1}{4}-\left\langle s_{\alpha }s_{\beta
}\right\rangle _{\alpha \neq \beta \in \mathcal{C}}\right] =\beta
g_{1}(\Sigma _{0})  \nonumber \\
&=&\beta \int_{-\infty }^{\infty }dyP(y)\frac{1}{[2\cosh {(\beta y/2)}]^{2}},
\label{susc}
\end{eqnarray}%
where we have introduced the charge susceptibility $\chi $, the analog of
the zero field cooled susceptibility in spin glasses. Expanding $g_{n}$ for $%
\Sigma _{0}/\beta \gg 1$, one can see that at low temperatures these two
equations only admit a solution if $\Sigma _{0}\sim \beta ^{3}$ and $P(y)$
takes the scaling form 
\begin{equation}
T^{-2}P(y\equiv zT)\rightarrow p(z)\quad \quad ({T\rightarrow 0})
\label{scaling}
\end{equation}%
with $p(z)\sim z^{2}$ for $z\gg 1$. This implies that the susceptibility
obeys the scaling $\chi \sim T^{2}$, and the (static) screening length
diverges at low temperatures as $l_{\mathrm{sc}}=(4\pi \chi )^{-1/2}\sim
T^{-1}$. Note that $\chi $ measures the charge response to a local potential
change when the particles on other sites are allowed to readjust to the
induced charge. Thus, it is associated with the thermodynamic local fields $%
y_{i}$ defined by $\left\langle s_{i}\right\rangle =m_{i}=\tanh (\beta
y_{i}/2)/2$. While we expect $\chi $ to control the hopping conductivity,
tunneling experiments~\cite{Cbgap} probe the system on very short time
scales, sampling the distribution $\widetilde{P}(h)$ of \emph{instantaneous}
local fields $h_{i}=\sum_{j}\mathcal{J}_{ij}s_{j}$. The %here 
\emph{thermal average} of these local fields, $\left\langle
h_{i}\right\rangle =\sum_{j}\mathcal{J}_{ij}m_{j}$, is related to the
thermodynamic field $y_{i}$ via a Thouless-Anderson-Palmer (TAP) equation~%
\cite{ThoulessAnderson77}, $\left\langle h_{i}\right\rangle
=y_{i}+\left\langle s_{i}\right\rangle h_{O}$, where the Onsager term 
\begin{equation}
h_{O}=\beta \int_{k}\frac{J_{k}^{2}}{\beta J_{k}+\Sigma _{0}}\approx 2\sqrt{%
\pi \beta /\Sigma _{0}}\approx 2\sqrt{\pi \chi }  \label{Onsager}
\end{equation}%
accounts for the extra polarizations induced by the presence of the charge $%
\left\langle s_{i}\right\rangle $. For the consistency with the locator
approximation, we have retained only terms corresponding to a local
self-energy. The deviation of the local field $h_{i}$ from its mean $%
\left\langle h_{i}\right\rangle $ is essentially a Gaussian variable with
width $h_{O}$. More precisely, the relation 
\begin{equation}
\widetilde{P}(h)=\int dyP(y)\frac{\cosh (\beta h/2)}{\cosh (\beta y/2)}\frac{%
e^{-\beta (h-y)^{2}/2h_{O}}}{\sqrt{2\pi h_{O}/\beta}e^{-\beta h_{O}/8}},
\label{Phinst}
\end{equation}%
holds~\cite{elsewhere}, which generalizes a known result for the SK-model~%
\cite{Thomsen86}. The tunneling density of states at zero bias then follows
from $\nu _{0}=\beta \int dh\widetilde{P}(h)[2\cosh (\beta h/2)]^{-2}$. Eq.~(%
\ref{Phinst}) implies that $\widetilde{P}(h)$ obeys a scaling analogous to
Eq.~(\ref{scaling}), and hence $\nu _{0}\sim T^{2}$. Generally, in order to
make quantitative predictions, one needs to know the functional form of the
field distributions. It turns out, however, that certain parameters are not
very sensitive to their details. It is convenient to assume a simple form $%
\overline{P}(\left\langle h\right\rangle )=\alpha (\left\langle
h\right\rangle ^{2}+\gamma T^{2})$ for the distribution of average fields,
obtain $P(y)$ via the TAP-equations and solve Eqs.~(\ref{margstab},\ref{susc}%
). This yields $\chi $, $\nu _{0}$ and $\alpha $ as slowly varying functions
of $\gamma $~\cite{SKestimeates}: $\alpha \approx 0.204-0.0067\gamma $, $%
\chi \approx (22.27-0.81\gamma )\,T^{2} $, $\nu _{0}\approx
(2.178-0.008\gamma )T^{2}$. The tunneling DOS $\nu _{0}$ is roughly an order
of magnitude smaller than the full susceptibility $\chi $, as is also
evident from the typical distributions shown in Fig.~\ref{fielddist}. This
agrees well with the experimental observation~\cite{multiparticle} that the
susceptibilities governing tunneling and hopping transport differ
significantly. 
The value of $\alpha $ should be compared to the Efros-Shklovskii prediction 
$\alpha _{ES}=3/\pi \approx 0.95$~\cite{efrosshklovskii7576} which is larger
than our estimate because their self-consistency argument imposes stability
only with respect to single electron hops. By contrast, our estimate
includes multiparticle constraints that decrease $\alpha $ below $\alpha
_{ES}$ in agreement with large-scale numerical simulations \cite{Moebius93}.

In conclusion, we have developed the locator approximation for Coulomb
glasses, allowing us to include multiparticle correlations. We have used
this formalism to provide evidence for a continuous glass transition below
which the Coulomb glass gets stuck in a marginally stable state, resulting
in subexponential relaxation dynamics and giving rise to the Coulomb gap. A
priori, the locator approximation is justified for large disorder. However,
as long as crystallization is prevented, a structural glass transition will
provide sufficient self-generated disorder, so that we expect our results to
hold at low temperatures even in the case of weak external disorder. We
verified \cite{elsewhere} that in this limit the local observables are still
determined by the contribution from large scales and reveal the
Efros-Shklovskii gap. Further, we found that the locator approximation gives
a significant decrease of the DOS with temperature already above $T_{c}$, in
agreement with numerics. Moreover, it predicts a discontinuous glass
transition at a scale of $T_{c}\approx 0.030$ which depends, however, on the
details of the cutoff at small scales. The validity of this prediction
remains thus unclear.

The locator approximation not only provides new insight into classical
Coulomb glasses, but also allows for quantitatively new predictions that go
beyond the single particle theory. It thus sets the stage for further
theoretical developments to understand the puzzles of correlated transport
and glassy relaxation of these systems. For instance, it allows one to study
the collective modes of the electrons which induce fluctuations in the local
electric fields~\cite{EGdynnum} and thus enhance the probability for
resonant tunneling. Experiments indicate that such a mechanism might provide
an alternative to phonon assisted tunneling, in particular at low
temperatures when phonons freeze out~\cite{eeconductivity}. Finally,
extensions of the formalism to include quantum effects may be envisioned.

We thank E. Abrahams, P. Chandra, V. Dobrosavljevi{\'{c}}, M. Feigel'man, G.
Kotliar, E. Lebanon, L. Leuzzi, Z. Ovadyahu, A. Silva, and especially B.
Altshuler (who participated in this work at the early stages) for
discussions. This work was supported by NSF grant DMR 0210575 and by grant
PBSK2 -10268/1 from the Swiss National Science Foundation.

\pagebreak \appendix

\section{Appendix: Justification of the locator approximation}

In order to justify the locator approximation we calculate the leading terms
ignored by the locator approximation and show that their effect on the
physical results is small in the limit $W\gg 1$. To calculate the higher
order terms we need a diagram technique. This is not convenient in the spin
representation of the original problem 
\begin{equation}
H[\{s_{i}\}]=\frac{1}{2}\sum_{i\neq j}s_{i}\mathcal{J}_{ij}s_{j}+%
\sum_{i}s_{i}\epsilon _{i},  \label{Hamiltonian2}
\end{equation}%
but turns out to be rather simple in terms of the fields $\phi _{i}$
conjugated to the spin variables $s_{i}$ that appear after we perform a
Hubbard-Stratonovich transformation: 
\begin{equation}  \label{Z}
Z=\sum_{\{s_{i}\}}e^{-\beta H[\{s_{i}\}]}=\int \prod_{i}d\phi
_{i}\sum_{\{s_{i}\}}e^{-\frac{1}{2}\sum_{i,j}\phi _{i}(\beta \mathcal{J}%
)_{ij}^{-1}\phi _{j}+\sum_{i}(\beta \epsilon _{i}+i\phi _{i})s_{i}}.
\end{equation}%
The field $i\beta^{-1}\phi_i$ can be thought of as the mean Coulomb
potential at site $i$ created by all other electrons \cite{Srinivasan71}.

In order to get rid of the disorder and to restore translational invariance,
we apply the replica trick and calculate $\left\langle Z^{n}\right\rangle
_{\epsilon }$, where $\left\langle {}\right\rangle _{\epsilon }$ denotes the
average over the random energies $\epsilon _{i}$. We assume the latter to be
independent local variables with distribution $P(\epsilon )$. The replicated
partition function, summed over spin degrees of freedom, reads 
\[
\left\langle Z^{n}\right\rangle _{\epsilon }=\left\langle \int
\prod_{i,a}d\phi _{i}^{a}e^{-\frac{1}{2}\sum_{i,j,a}\phi _{i}^{a}(\beta 
\mathcal{J})_{ij}^{-1}\phi _{j}^{a}+\sum_{i,a}\ln [\cosh (\frac{\beta
\epsilon _{i}+i\phi _{i}^{a}}{2})]}\right\rangle _{\epsilon }.
\]%
We expand the local terms $\ln [\cosh (\frac{\beta \epsilon _{i}+i\phi
_{i}^{a}}{2})]$ with respect to $\phi _{i}^{a}$ and evaluate the disorder
average by a cumulant expansion, 
\begin{eqnarray}
\left\langle Z^{n}\right\rangle _{\epsilon } &=&\int \prod_{i,a}d\phi
_{i}^{a}\exp \left[ -\frac{1}{2}\sum_{i,j,a}\phi _{i}^{a}(\beta \mathcal{J}%
)_{ij}^{-1}\phi _{j}^{a}-\frac{1}{2}\frac{1}{\beta \widetilde{W}}%
\sum_{i,a}(\phi _{i}^{a})^{2}-\frac{1}{2}\zeta \sum_{i,a,b}\phi _{i}^{a}\phi
_{i}^{b}\right.   \nonumber  \label{Zn} \\
&&\quad \quad \quad \quad \quad \quad \quad \quad \left. +\frac{\eta }{4!}%
\sum_{i,a}(\phi _{i}^{a})^{4}+\frac{1}{2}\frac{\lambda }{(2!)^{2}}%
\sum_{i,a,b}(\phi _{i}^{a}\phi _{i}^{b})^{2}+\dots \right] 
\end{eqnarray}%
where $\frac{1}{\beta \widetilde{W}},\zeta ,\eta ,\lambda $ denote vertex
coefficients: 
\begin{eqnarray}
\frac{1}{\beta \widetilde{W}} &=&\left\langle \frac{1}{[2\cosh (\beta
\epsilon /2)]^{2}}\right\rangle _{\epsilon }\approx \frac{1}{\sqrt{2\pi }%
\beta W},  \nonumber \\
\zeta  &=&\left\langle {[\tanh ^{2}(\beta \epsilon /2)/2]^{2}}\right\rangle
_{\epsilon }\approx \frac{1}{4}(1-\frac{1}{\sqrt{2\pi }\beta W}),  \nonumber
\\
\eta  &=&\left\langle \frac{d^{2}}{d(\beta \epsilon )^{2}}\left( \frac{1}{%
[2\cosh (\beta \epsilon /2)]^{2}}\right) \right\rangle _{\epsilon }\approx 
\frac{1}{\sqrt{2\pi }(\beta W)^{3}},  \nonumber \\
\lambda  &=&\left\langle \frac{1}{[2\cosh (\beta \epsilon /2)]^{4}}%
\right\rangle _{\epsilon }-\left\langle \frac{1}{[2\cosh (\beta \epsilon
/2)]^{2}}\right\rangle _{\epsilon }^{2}\approx \frac{1}{6\sqrt{2\pi }\beta W}%
.  \nonumber  \label{Znexpand}
\end{eqnarray}

Here we assumed a symmetric distribution of $\epsilon _{i}$ and dropped
the zeroth order term in (\ref{Zn}) since it is irrelevant in the replica
limit $n\rightarrow 0$. The last equality in each equation gives the leading
order in $1/\beta W$ for Gaussian disorder. Note that typically vertex
coefficients scale as $\sim 1/(\beta \widetilde{W})$, except for the
coefficients $c_{m}$ corresponding to the replica conserving vertices $(\phi
_{i}^{a})^{m}$. The latter are suppressed by additional powers of $\beta 
\widetilde{W}$, $c_{m}\sim 1/(\beta \widetilde{W})^{m-1}$ because they
correspond to disorder averages of exact derivatives of $1/\cosh ^{2}[\beta
\epsilon /2]$. To establish the connection with the locator approximation,
notice that in the leading order in $1/\beta W$ the quantity $\widetilde{W}$
coincides with the  diagonal part of the self energy  $\Sigma _{0}$ of the
replica symmetric theory. 

A systematic diagrammtic expansion is now obtained by taking $%
[G_{0}^{-1}]_{ij}^{ab}=\left\{ (\beta \mathcal{J})_{ij}^{-1}+\delta
_{ij}/\beta \widetilde{W}\right\} \delta ^{ab}+\zeta \delta _{ij}\mathcal{I}%
^{ab}$ as bare propagator and treating all other terms in the cumulant
expansion as interactions. In Fourier space, we have 
\begin{eqnarray}
G_{0}^{ab}(k) &=&g_{0}(k)\delta ^{ab}+g_{1}\mathcal{I}^{ab},  \label{G_0} \\
g_{0}(k) &=&\frac{\beta }{\frac{k^{2}}{4\pi }+\frac{1}{\widetilde{W}}}, 
\nonumber \\
g_{1}(k) &=&-\zeta g_{0}^{2}(k).  \nonumber
\end{eqnarray}

The leading non-local contribution, $\Sigma ^{(1)}$, to the replica diagonal
part of the self-energy is a tripled propagator line, connecting two
replica-conserving 4-vertices. Its $k$-dependent part evaluates to

\begin{eqnarray}  \label{fullSigma1}
\Sigma _{ab}^{(1)}(k)-\Sigma _{ab}^{(1)}(0) &=&\delta _{ab}\eta^{2}\int
g_{0}^{3}(r)(\exp (ikr)-1)d^{3}r  \label{Sigma1} \\
&=&\eta^2\cdot 4\pi\beta^3\delta_{ab}\left\{ 1-\frac{3}{r_0k} \arctan \left( 
\frac{r_0k}{3}\right) -\frac{1}{2}\ln \left[1+\left( \frac{r_0k}{3}\right)
^{2}\right] \right\},  \nonumber
\end{eqnarray}
where $r_0=(\widetilde{W}/4\pi)^{1/2}$. The local part is of the order of 
\begin{eqnarray}
\Sigma _{ab}^{(1)}(0) &\approx&\delta _{ab}\eta^2\cdot
4\pi\beta^3\ln(r_0)\sim \delta _{ab}\frac{\ln(\widetilde{W})}{\beta^3%
\widetilde{W}^6}.  \nonumber
\end{eqnarray}

As we will show below, the glass physics is dominated by long scales, $%
k\lesssim 1/r_{0}$. For these momenta one can approximate the momentum
dependent part of the self energy by 
\[
\Sigma _{ab}^{(1)}(k)-\Sigma _{ab}^{(1)}(0)\approx -\frac{4\pi }{54}\eta
^{2}\beta ^{3}\left( r_{0}k\right) ^{2}+O\left( \left( r_{0}k\right)
^{4}\right) .
\]%
which results in an insignificant renormalization of the Coulomb interaction
at temperatures of the order of the glass transition ($\beta _{c}\sim W^{1/2}
$). In this regime we can thus safely approximate the self-energy by its
local part $\Sigma ^{(1)}(0)$.

At very low temperatures the density of states at low energies is
suppressed, this reduces the screening of the electric field and enhances
the interactions. In this regime, the renormalizations might become more
important. To estimate this effect we replace the screening part of the bare
propagator, $1/\widetilde{W}=\beta \langle \lbrack 2\cosh (\beta \epsilon
/2]^{-2}\rangle _{\epsilon }$ by the average thermodynamic susceptibility $%
\chi =\beta \int dyP(y)[2\cosh (\beta y/2]^{-2}\sim 1/\beta ^{2}$. This
gives rise to the bare propagator $g_{0}(k)=\beta /(k^{2}/4\pi +\chi )$ with
the longer screening length  $r_{0}\rightarrow (4\pi \chi )^{-1/2}\sim \beta 
$. Further, as a consequence of the broken replica symmetry, the replica
mixing vertices might also contribute to the renormalization of the replica
diagonal part of the Greens function. At low temperatures the vertex
coefficients generically scale as $1/\beta ^{3}$. This might increase the
importance of the non-local corrections and even make them marginally
relevant below some temperature $T^{\ast }$. By continuity we must have $%
T^{\ast }\ll T_{c}$. We do not know any physical argument that supports the
appearance of the second temperature (energy)\ scale and it seems unlikelt
to us that it happens. However, in order to check that non-local
contributions remain parametrically small at all temperatures one needs to
calculate them  against the background of the full replica symmetry breaking
solution. 

%\textbf{Rather drop this:} Even at smaller scales this approximation does not lead to significant corrections to the full Greens function, $G(k)=G_{0}(k)/(1-\Sigma (k)G_{0}(k))$, since the product
%\begin{equation}
%\Sigma (k)G_{0}(k)\sim \frac{\ln(\beta)}{(\beta k)^2},\quad \quad fro \quad k>1/\beta   \label{locality}
%\end{equation}%
%is (numerically) small compared to $1$ for $k>T$. 

Incorporating the local term $\delta _{ij}\delta ^{ab}/\widetilde{W}$ of the
bare propagator into the self-energy $\Sigma _{\phi }$ we get the full
Greens function 
\begin{equation}
\left\langle \phi _{i}^{a}\phi _{j}^{b}\right\rangle _{c}=[G^{-1}]_{ij}^{ab}=
\left[ \frac{1}{(\beta \mathcal{J})^{-1}-\Sigma _{\phi }}\right] _{ij}^{ab}.
\end{equation}

Before discussing the physical interpretation of this quantity in more
detail we mention that the spin-spin correlation function follows easily via
partial integration in (\ref{Z}) with respect to the conjugated fields, 
\begin{equation}  \label{correlator}
\left\langle s_i^as_j^b\right\rangle_c = (\beta \mathcal{J})^{-1}_{ij}
-\sum_{l,m}(\beta\mathcal{J})^{-1}_{il}\left\langle\phi^a_l\phi^b_m\right%
\rangle_c (\beta\mathcal{J})^{-1}_{mj}=\left[\frac{1}{\beta \mathcal{J}%
-\Sigma}\right]^{ab}_{ij},
\end{equation}
where $\Sigma =\Sigma^{-1}_\phi$.

The locality of the self-energy suggests the mapping to an effective single
site model, for which the same self-energy is assumed. For self-consistency
we then have to require that the average local spin-spin correlators be the
same. 
\begin{equation}  \label{correlatoronsite}
\left\langle s^a s^b\right\rangle_c = \left[\frac{1}{\mathcal{B}-\Sigma}%
\right]^{ab}=\frac{1}{V}Tr \left\langle s_i^a s_j^b\right\rangle_c = \frac{1%
}{V}Tr \left[\frac{1}{\beta \mathcal{J}-\Sigma}\right]^{ab}.
\end{equation}

The equation (\ref{correlator}) for spin-spin correlator  is a central
element in the mapping to a single site-model. Physically, this expression
and the locator approximation in general, is based on the assumption that a
typical spin interacts with many thermally active neighbors. The
calculations above show that this assumption is satisfied by a typical spin
that contributes to the density of states and other thermodynamic
properties. This, however, does not exclude the presence of ther types of
low-energy excitations. For instance, an occupied and a nearby empty site
may form a strongly coupled dipole, in which the electron can hop to the
empty site at very low energy expense. Such pairs of sites are strongly
coupled, and are not described by the locator approximation. The small
effect of non-local corrections evaluated here physically mean that such
dipoles renormalize weakly the dielectric susceptibility without interfering
strongly in thermodynamics of carrier sites. The effect of these dipoles on
other properties remain an open question. 

\subsection{Fluctuating correlations and criticality}

The replica diagonal part of the Green function (\ref{averagescreening}) is
simply the disorder average (or equivalently, the average over sites at
fixed distance) of the connected correlation function. Assuming a local
self-energy, we explicitly obtain 
\begin{equation}  \label{averagescreening}
\left\langle \phi_i^a\phi_j^a\right\rangle_c=\overline{\left\langle
\phi_i\phi_j\right\rangle_c}= \beta \frac{e^{-r/r_0}}{r}
\end{equation}
where the overbar denotes the disorder average, and $r_0=\sqrt{%
\Sigma_0/(4\pi\beta)}$ ($=(\widetilde{W}/4\pi)^{1/2}$ in the limit of large $%
W$ and at temperatures above $T_c$). This expression may be misleading since
it seems to suggest that potentials are screened on average on a distance $%
r_0$. However, we have to bear in mind that (\ref{averagescreening})
describes only the \emph{disorder averaged} correlations. To study the
fluctuations around this average, we have to calculate $\overline{\left\langle \phi_i\phi_j\right\rangle_c^2}-\overline{\left\langle \phi_i\phi_j\right\rangle_c}^2$. Within the replica formalism the first term
can be calculated as 
\begin{equation}
\label{fourpoint}
\overline{\left\langle \phi_i\phi_j\right\rangle_c^2}=\left\langle (\phi^a_i\phi^a_j -\phi^a_i\phi^c_j)-(\phi^b_i\phi^b_j-\phi^b_i\phi^d_j)\right\rangle\, ,
\end{equation}
$a,b,c$ and $d$ being all distinct replicas that see the same disorder
configuration. (In the glass phase one has to choose all indices in the same
innermost block, which imposes that the two replicas belong to the same
metastable state.)

\begin{figure}[tbp]
\resizebox{7.5 cm}{!}{\includegraphics{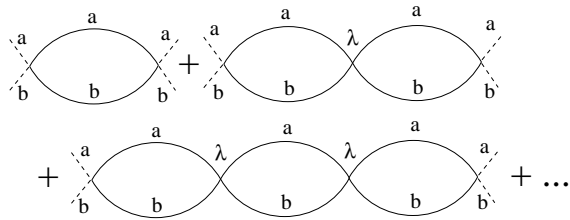}}
\caption{Leading diagrams for the computation of the four point function $%
\overline{\left\langle \protect\phi _{i}\protect\phi _{j}\right\rangle ^{2}}$%
.}
\label{figfourpoint}
\end{figure}

This four point function can be evaluated by the diagram technique, the
leading diagrams being shown in Fig.~\ref{figfourpoint}. 
%We focus on the
%replicon mode defined as the 
The correlation function (\ref{fourpoint}) is given by the coefficient, $C$, of $\delta _{ac}\delta _{bd}$ of
the full expression for $\left\langle \phi _{i}^{a}\phi _{j}^{a}\phi
_{i}^{c}\phi _{j}^{d}\right\rangle$. It can easily be calculated by resumming a
geometric series, which yields 
\begin{equation}
C(r)=\int d^{3}re^{ikr}\overline{\left\langle \phi (0)\phi (r)\right\rangle
^{2}}=\Sigma ^{(2)}(k)\left[ 1+\lambda \Sigma ^{(2)}(k)+\left( \lambda
\Sigma ^{(2)}(k)\right) ^{2}+\dots \right] =\frac{1}{1/\Sigma
^{(2)}(k)-\lambda }.  \label{4pointseries}
\end{equation}%
Here, $\Sigma ^{(2)}(k)$ denotes the insertion of a doubled propagator line, 
\[
\Sigma ^{(2)}(k)=\int g_{0}^{2}(r)e^{ikr}d^{3}r=2\pi r_{0}\left( 1-\frac{%
(r_{0}k)^{2}}{12}+O((r_{0}k)^{4})\right) .
\]%
Using the expressions at high temperature in the large $W$-limit, one easily
verifies that at the critical temperature $T_{c}=W^{-1/2}/[6(2/\pi )^{1/4}]$
(cf. Eq. (\ref{RSinstab})), the $k$-independent terms in (\ref{4pointseries}%
) cancel, leading to a power law decay of the potential correlations, 
\begin{equation}
\overline{\left\langle (\beta ^{-1}\phi (0))(\beta ^{-1}\phi
(r))\right\rangle _{T=T_{c}}^{2}}\sim \frac{1}{\widetilde{W}^{2}}\frac{\beta _{c}}{r}%
.  \label{critcorr}
\end{equation}%
We separated a prefactor $1/\widetilde{W}^{2}$ which has the interpretation of the
probability that both sites at $0$ and $r$ are thermally active. 

Note that the critical temperature is exactly the same as we found by
mapping to a single-site problem. In the original lattice, the replicon
instability receives a natural interpretation: The correlations in the
Coulomb potential created by the electron configuration become critical at $%
T_{c}$ where they only decay with a power law. Below $T_{c}$, the phase
space splits into an exponential number of metastable states (ergodic
components) each of which is characterized by the finite expectation values $%
\psi _{i}$ of the conjugated potential fields $\psi _{i}=i\beta ^{-1}\langle
\phi _{i}\rangle $. \textbf{\ } From Eq.~(\ref{Z}) we see that within such a
metastable state we should consider $\epsilon _{i}+\psi _{i}$ as the
effective field on the site $i$ to be used to calculate the vertex
coefficients. Anticipating an Efros-Shklovskii-type distribution for those
effective fields then suggests that the vertex coefficients averaged over
sites, scale as $1/\beta ^{3}$ as mentioned above. In particular, the mass
term of the propagator will be replaced by the average susceptibility
corresponding to the distribution of effective fields $\epsilon _{i}+\psi
_{i}$.

At this point the diagram technique in the original lattice becomes very
complicated since translational invariance is spontaneously broken by the
emergence of spontaneous expectation values. The advantage of the mapping to
a single site problem is now obvious: The replica structure implicitly takes
the statistics of a spin's environment into account, and allows, e.g., for a
self-consistent determination of the distribution of local fields.

\subsection{Spin-spin correlations and inhomogeneous charge response}

The criticality of the correlations of potential fields translates directly
to the criticality of spins. The fact that the glass phase is marginally
stable has the simple interpretation that these correlations remain long
range (power law) throughout the low temperature phase. This self-organized
criticality has an important consequence: One can show that by imposing the
spin on site $i$ to take a definite value one induces a polarization at site 
$j$ which is proportional to $\langle s_is_j\rangle_c$. The above results
imply that this response of the system in the glass phase decays only as a
power law with distance, i.e., screening is almost absent. Furthermore, we
note that the polarization induced on different sites do not have a definite
sign. It should be clear from these considerations that typical spins
interact with a large number of effective neighbors, which in turn
self-consistently confirms the initial assumption suggesting the locator
approximation.

\section{Appendix: The replicon instability}

Within the effective single-site model, the glass transition (\ref{RSinstab}%
) and the marginal stability condition (\ref{margstab}) both derive from the
vanishing of the eigenvalue of the Hessian $\partial^2 F/\partial \mathcal{B}%
^2$ in the replicon mode. The variation of the spin-part in (\ref{F(B)}) is
standard and yields 
\begin{equation}  \label{spinpart}
\delta^2 \beta F_{spin} = -tr (\delta \mathcal{B}^2) \int_{-\infty }^{\infty
}dyP(y)\frac{1}{[2\cosh (\beta y/2)]^{4}}.
\end{equation}

When varying $U(\mathcal{B})$, (\ref{U(B)}), we need to take into account
that $\Sigma$ is a function of $\mathcal{B}$. The selfconsistency condition 
\begin{eqnarray}
\frac{1}{\mathcal{B}-\Sigma }=g_1(-\Sigma)
\end{eqnarray}%
imposes, upon variation in the replicon mode, that $(\delta\mathcal{B}%
-\delta\Sigma)g_1^2(\Sigma_0)=-g_2(\Sigma_0)\delta\Sigma$. The variation of $%
U(\mathcal{B})$ thus evaluates to 
\begin{eqnarray}  \label{Upart}
\delta^2 U(\mathcal{B})=-g_1^2(\Sigma_0)tr(\delta\mathcal{B}%
-\delta\Sigma)^2+g_2(\Sigma_0)tr(\delta\Sigma)^2= \frac{tr(\delta \mathcal{B}%
^2)}{g_{1}^{-2}(\Sigma _{0})-g_{2}^{-1}(\Sigma _{0})},
\end{eqnarray}
which together with (\ref{spinpart}) leads to (\ref{RSinstab}) and (\ref%
{margstab}).

\section{Appendix: Generalized Onsager term}

We have shown above that the spin $s_0$ at site $0$ polarizes its
environment in a large spatial region. As inis well-known from the
TAP-approach to spin glasses, in order to obtain the thermodynamic field $%
y_0 =2\beta^{-1} \tanh^{-1}(2 m_0)$, the back reaction of this polarization
on the spin itself has to be subtracted from the thermally averaged field $%
\left\langle h_0 \right\rangle=-\sum_{j\neq 0}\mathcal{J}_{0j}m_j$, 
\begin{equation}  \label{TAPrel}
y_0= \left\langle h_0 \right\rangle_{s=s_0} - s_0 h_O,
\end{equation}
where $h_O$ is the famous Onsager back reaction. The usual term $h_O=
\sum_{j} \mathcal{J}^2_{ij} \chi_{j}$, familiar from spin glasses, has to be
generalized to the case of Coulomb glasses where this expression is clearly
divergent. Indeed, to obtain a finite response we have to sum up all higher
order polarizations, 
\begin{equation}  \label{generalizedOnsager2}
h_O= \sum_{j_1} \mathcal{J}_{ij_1} \chi_{j_1}\mathcal{J}_{j_1i}-\sum_{j_1,
j_2} \mathcal{J}_{ij_1} \chi_{j_1}\mathcal{J}_{j_1j_2}\chi_{j_2}\mathcal{J}%
_{j_2i} +\dots,
\end{equation}
their alternating sign reflecting the antiferromagnetic nature of the
Coulomb interactions \cite{Srinivasan71}. Approximating the on-site
susceptibilities $\chi_j$ by their average $\chi$ (which is justified since
we average over a very large number of spins), we may perform the sum 
\begin{equation}  \label{generalizedOnsager}
h_O= Tr \frac{\mathcal{J}^2}{\chi^{-1}+\mathcal{J}}\approx \beta\, Tr \frac{%
\mathcal{J}^2}{\beta\mathcal{J}+\Sigma_0}\approx 2\pi\sqrt{\Sigma_0}\sim T.
\end{equation}
The last approximations are valid at low temperatures.

The distribution of instantaneous fields can be obtained from 
\begin{eqnarray}
P(h,s)&=&\frac{1}{V}\sum_i\int \frac{d\lambda}{2\pi}\left\langle e^{i\lambda
h-i\lambda\sum_j \mathcal{J}_{ij}s_j} \delta_{s_i s}\right\rangle  \nonumber
\\
&=&\frac{1}{V}\sum_i\int \frac{d\lambda}{2\pi} e^{i\lambda h-i\lambda\sum_j
\left\langle \mathcal{J}_{ij}s_j\right\rangle_{s_i=s} -\lambda^2/2\sum_{j,k}%
\mathcal{J}_{ij}\left\langle s_j s_k\right\rangle_{c,s_i=s} \mathcal{J}%
_{ki}}.
\end{eqnarray}
We have only retained the first two cumulants, to be consistent within the
locator approximation. From the generalized TAP-equations (\ref{TAPrel}) we
may identify the first cumulant as 
\begin{equation}
\left\langle h_{i}\right\rangle_{s_i=s} = y_i+s h_O,  \label{TAPeq}
\end{equation}%
The second cumulant is almost insensitive to the value of the spin at site $%
i $, and evaluates to 
\begin{eqnarray}
\sum_{j,k}\mathcal{J}_{ij}\left\langle s_j s_k\right\rangle_{c} \mathcal{J}%
_{ki}=\left[\mathcal{J}\frac{1}{\beta\mathcal{J}+\Sigma_0}\mathcal{J}\right]%
_{ii}\approx Tr\left[\frac{\mathcal{J}^2}{\beta\mathcal{J}+\Sigma_0}\right]=
\beta^{-1} h_O.
\end{eqnarray}
in the locator approximation.

To carry out the site average, we have to weight the pairs $(y,s)$,
according to their joint probability density $P(y)\exp(\beta y
s)/(2\cosh(\beta y/2)$, 
\begin{equation}
P(h,s)=\int dy P(y) \left[\frac{1}{2\cosh(\beta y/2)}\int \frac{d\lambda}{%
2\pi} e^{s\beta y}e^{i\lambda (h-y-sH(\beta))}e^{-\lambda^2 H(\beta)/2\beta}%
\right].
\end{equation}
Performing the $\lambda$-integral we find 
\begin{equation}
P(h,s)=\int dy P(y) \frac{e^{s \beta y}}{2\cosh(\beta y/2)} \frac{\exp[%
-\beta(h-y-sh_O)^2/2h_O]}{[2\pi h_O/\beta]^{1/2}},
\end{equation}
and summing over $s$ we find the local field distribution 
\begin{equation}
P(h)=\int dy P(y) \frac{\cosh(\beta h/2)}{\cosh(\beta y/2)} \frac{\exp[%
-\beta(h-y)^2/2h_O]}{[2\pi h_O/\beta]^{1/2}}\exp(\beta h_O/8).
\end{equation}

\end{document}